\documentclass[aps,prx,twocolumn,superscriptaddress,notitlepage,nobibnotes,nofootinbib]{revtex4-1}
\usepackage[colorlinks=true, citecolor=magenta, linkcolor=blue, urlcolor=blue]{hyperref}
\usepackage{graphicx}
\usepackage{amsmath}
\usepackage{amssymb}
\usepackage{amsfonts}
\usepackage{soul}
\usepackage{xcolor}
\usepackage{url,physics,bm,float}
\allowdisplaybreaks

\begin{document}

\title{Microscopic theory of current-induced skyrmion transport and its application in disordered spin textures} 
\author{Emil \"Ostberg}
\email{emil.ostberg@teorfys.lu.se}
\affiliation{Department of Physics, Division of Mathematical Physics, Lund University, 22100 Lund, Sweden}
\author{Emil Vi\~nas Bostr\"om}
\email{emil.bostrom@mpsd.mpg.de}
\affiliation{Max Planck Institute for the Structure and Dynamics of Matter, Luruper Chaussee 149, 22761 Hamburg, Germany}
\affiliation{Nano-Bio Spectroscopy Group, Departamento de Fisica de Materiales, Universidad del Pais Vasco, 20018 San Sebastian, Spain}
\author{Claudio Verdozzi}
\email{claudio.verdozzi@teorfys.lu.se}
\affiliation{Department of Physics, Division of Mathematical Physics and ETSF, Lund University, 22100 Lund, Sweden}

\begin{abstract}
\noindent Magnetic skyrmions hold great promise for realizing compact and stable memory devices that can be manipulated at very low energy costs via electronic current densities. In this work, we extend a recently introduced method to describe classical skyrmion textures coupled to dynamical itinerant electrons. In this scheme, the electron dynamics is described via nonequilibrium Green's functions (NEGF) within the generalized Kadanoff-Baym ansatz, and the classical spins are treated via the Landau-Lifshitz-Gilbert equation. The framework is here extended to open systems, by the introduction of a non-interacting approximation to the collision integral of NEGF. This, in turn, allows us to perform computations of the real-time response of skyrmions to electronic currents in large quantum systems coupled to electronic reservoirs, which exhibit a linear scaling in the number of time steps. We use this approach to investigate how electronic spin currents and dilute spin disorder affects skyrmion transport and the skyrmion Hall drift. Our results show that the skyrmion dynamics is sensitive to the specific form of spin disorder, such that different disorder configurations leads to qualitatively different skyrmion trajectories for the same applied bias. This sensitivity arises from the local spin dynamics around the magnetic impurities, a feature that is expected not to be well captured by phenomenological or spin-only descriptions.
At the same time, our findings illustrate the potential of engineering microscopic impurity patterns to steer skyrmion trajectories.
\end{abstract}

\maketitle


\section{Introduction}\label{Sec_Intro} 
Technological progress often stems from discovery and exploitation of new materials and forms of energy. While self-evident, this
paradigm has recently undergone criticism and revision, due to mounting awareness of the negative impact that indiscriminate technological development has on environment and climate. This is also true for electronics: it has become clear that  production, use and casual disposal of electronic devices can lead to a sharp increase in energy consumption, waste, and greenhouse effects~\cite{Waste}.  Thus, together with a steady increase in the use of high performance technology, there is a need for novel electronics with reduced dimensionality, large integration and low energy consumption~\cite{Green}. 

Pursuing on equal footing these two directives is the core aim of spintronics~\cite{Spintronics1}: magnetic excitations allow for less energy-intensive ways of storing and processing digital information, and thus devices based on magnetic materials and phenomena offer an attractive alternative to conventional electronics. For a long time, fundamental and applied research in magnetism was largely concerned with macroscopic samples, and primarily with simple magnetic orders, such as ferro- and antiferromagnets. However, more recently, it has been possible to experimentally realize magnetic systems with non-trivial magnetic textures, creating unprecedented possibilities for spintronic applications~\cite{Spintronics2,Parkin2015}.

A notable example in this respect is provided by magnetic skyrmions \cite{Mertig2021}. These are topologically nontrivial spin textures stabilized by a competition of exchange, Dzyaloshinskii-Moriya interactions (DMIs) and magnetic anisotropies~\cite{Skyrmionics}. With their compact size, topological protection, and non-intensive energy requirements for manipulation, skyrmions are of great potential interest to realize race track memories~\cite{Sampaio2013,Nagaosa,Romming2015} and quantum computation devices~\cite{Yang16,Chauwin19,Mirlin19,Psaroudaki2021}. This, however, requires efficient ways of writing, deleting, and manipulating skyrmions on short time scales and with high spatial precision, via electronic spin currents. 

\begin{figure}
\begin{center}
 \includegraphics[width=0.5\textwidth]{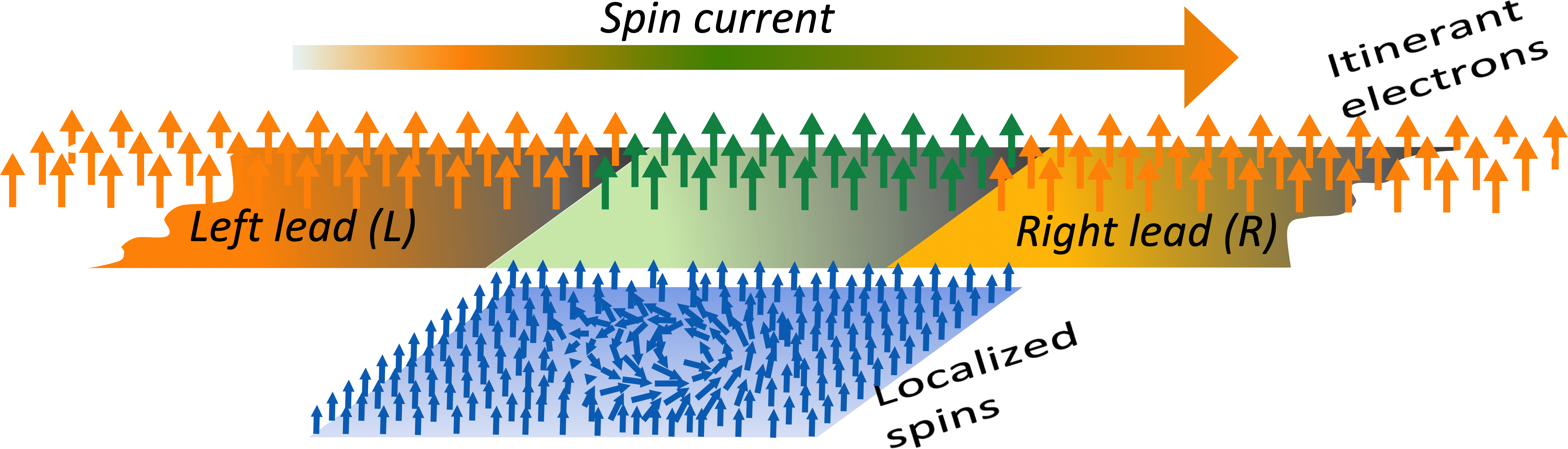}
 \end{center}
 \caption{{\bf A sketch of the composite spin-electron system.} A central region consisting of itinerant electrons interacting with localized spins is in contact with two-dimensional leads. By applying a spin dependent bias in the leads an electronic current is generated in the central system giving rise to skyrmion motion.}
 \label{fig:Scheme}
\end{figure} 

Here we describe a scheme to perform large scale simulations of the intertwined dynamics of interacting and open spin-electron systems (Fig.~\ref{fig:Scheme}). Our method is an extension of the approach introduced in Refs.~\cite{LeadsPaper,LaserPaper}, and amounts to propagating the equation of motion for the electronic spin-dependent single-particle density matrix together with the Landau-Lifshitz-Gilbert equation for the classical spins. The former equation can be derived from the general theory of non-equilibrium Green's functions using the so-called generalized Kadanoff-Baym ansatz (GKBA)~\cite{Lipavsky86,Hermanns12,Hermanns14,Latini14,Bostrom18,Hopjan18,KarlssonHopjan18,Kalvova18,Joost20,Karlsson21,Pavlyukh22a,Pavlyukh22b,BonitzWBL23,Tuovinen23}, and therefore allows to systematically introduce the effects of electron-electron interactions via diagrammatic many-body perturbation theory. In what follows, we apply this scheme to study current-induced skyrmion motion, fully accounting for the dynamics of the itinerant electrons resulting from an applied bias. In agreement with phenomenological theories~\cite{Iwasaki13}, we find that for a clean sample skyrmions are pinned below a critical spin current density $I_0$, after which the velocity is a linear function of the current $I - I_0$.  The situation is found to be qualitatively different in presence of (dilute) spin disorder, where the skyrmion motion is strongly dependent on the location, size and form of the disorder configuration (for previous work on the role of disorder, see e.g.\cite{Thorwart2021}). This provides a clear indication that treatments based on the standard Thiele or Landau-Lifshitz-Gilbert (LLG) equation are not always adequate, and that the dynamics of the electrons must be explicitly taken into account.

The manuscript is structured as follows: In Sec.~\ref{sec:review} we briefly review previous approaches to coupled spin-electron dynamics. In Sec.~\ref{sec:system_hamiltonian} we introduce the system and Hamiltonian to be considered, and discuss the coupling between the itinerant electrons and external reservoirs. The spin and electron equations of motion are presented in Sec.~\ref{sec:eom} and Sec.~\ref{sec:negf}, and Sec.~\ref{sec:awbl} introduces an {\it approximate wide band limit} (AWBL) as a numerically efficient way to propagate the equations of motion in presence of large central regions connected to external reservoirs. In Sec.~\ref{sec:equilibrium} we present some observables used to interrogate the skyrmion content of the spin configuration. In Sec.~\ref{sec:current_motion} we use the AWBL to investigate skyrmion motion induced by current densities in the itinerant electron system, and in Sec.~\ref{sec:disorder} we consider skyrmion motion in presence of magnetic disorder. Finally, in Sec.~\ref{sec:discussion} we conclude with a discussion of experimental signatures and possible material platforms for which our results are of relevance.


\section{Review of previous approaches}\label{sec:review}
Magnetic skyrmions are made up of localized magnetic moments, typically arising from a large Hund's coupling $J$ favoring a high-spin state the $d$- or $f$-orbitals of the magnetic ion. Therefore, from a microscopic perspective, it is natural to expect that a quantum description of the magnetic structure should be necessary \footnote{For current work in this direction, see e.g. \cite{Haller22}}. However, since magnetic moments in typical skyrmion materials (consisting of transition metal ions such as Fe, Co and Mn) are of large magnitude, spin fluctuations are suppressed and a classical approximation usually works well.

In most cases skyrmion textures are large compared to the underlying lattice constant. More precisely, the ratio of the electronic hopping $t$ and the strength $\alpha$ of the spin-orbit interaction is typically of the order $\alpha/t \sim 0.01 - 0.1$, resulting in a spin spiral wavelength $\lambda \sim 10 - 100$ nm~\cite{Nagaosa,Experimental_review}. In certain cases however, such as at an interface between metallic thin films and a material with large spin-orbit coupling, the effective Dzyaloshinksii-Moriya interaction can be significantly enhanced and lead to skyrmions with radii on the order of a few nm~\cite{Heinze11}. For skyrmions of large sizes it is common to take a continuum limit of the microscopic spin Hamiltonian, resulting in a magnetic energy functional that can be minimized with micromagnetic methods.
As a result, calculations for lattice skyrmions of realistic size (up to about $10$ nm in radius in two-dimensional systems) are  usually  based on an atomistic description of classical spins, with external fields such electromagnetic radiation or electronic current densities included as non-dynamics variables. For example, the motion of skyrmions in response to an external current density can been described via an effective equation for the skyrmion center of mass, the so-called Thiele equation~\cite{Thiele73}, assuming that the form of the skyrmion is rigid. In the Thiele equation, electrons enter only via the external current density, taken from a static solution of the macroscopic Maxwell equations. A more detailed description often considered in the literature is to obtain the individual spin dynamics from the Landau-Lifshitz-Gilbert (LLG) equation~\cite{Iwasaki13}, and to include the effects of electrons and external fields via a generalized force~\cite{Nikolic1,Nikolic2}. The LLG equation, with itinerant electrons included implicitly, has provided important insights into skyrmion behavior in a large range of materials, and its use is widespread. Furthermore, other approaches have been introduced, besides the LLG, that go beyond the semiclassical Thiele's description to study the dynamics of skyrmions \cite{
PhysRevLett.127.047203, PhysRevB.107.224418, PhysRevB.107.174418, PhysRevB.108.064427}.

Indeed, there are many cases where neglecting the explicit dynamics of the electrons severely hinders a more detailed understanding, and possibly even the development of novel physical ideas and technological opportunities in skyrmionics. 

The importance of explicitly accounting for itinerant electrons in the description of skyrmion dynamics was originally pointed out in Ref.~\cite{LeadsPaper}, where a small two-dimensional spin texture containing a single skyrmion was made to interact with a nanowire carrying a time-dependent current. Since then, the significance of including electronic degrees of freedom in the dynamical description of skyrmions (as well as more general spin textures), has been addressed in several contexts~ \cite{LaserPaper,bluegel1,Sinova,Bluegel2,Nikolic1,Nikolic2}. For example, it was recently demonstrated~\cite{LaserPaper} that explicitly accounting for the dynamics of itinerant electrons can be of crucial importance in capturing the interaction between skyrmions and laser light. In particular, such a treatment shows that photo-generation of skyrmions resulting from laser excitation can occur on much shorter timescales than previously thought. 

Similarly, electrons and currents are expected to play a key role in skyrmion transport phenomena both for clean samples and when disorder is present. However, to investigate such dynamics requires to overcome an additional hurdle on the methodological side, namely that it is necessary to simultaneously treat large system sizes and extended time scales within an open quantum system framework. In the remainder of this manuscript, we show how such a framework can be constructed.


\section{System and Hamiltonian} \label{sec:system_hamiltonian}
\begin{figure}
\begin{center}
 \includegraphics[width=0.5\textwidth]{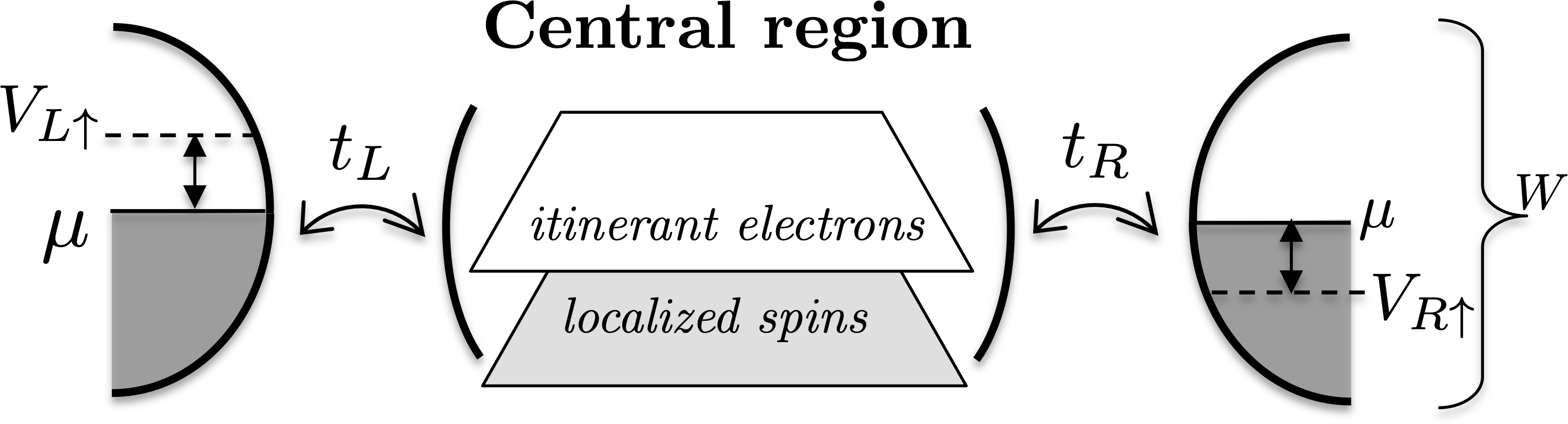}
 \end{center}
 \caption{{\bf Schematic illustration of the transport setup and the relevant parameters of the system-reservoir Hamiltonian}. The leads have a common bandwidth $W$ and chemical potential $\mu$, and are connected to a central region of $20 \times 40$ sites comprising both electrons and localised spins via a nearest-neighbour tunneling amplitude $t_{L/R}$ at the left/right edge. To initiate the skyrmion dynamics, a spin-dependent bias ($V_{L\uparrow}=-V_{R\uparrow}$ and $V_{L\downarrow}=V_{L\downarrow}=0$) is applied in the leads.}
 \label{fig:SchemeEmil}
\end{figure}

We consider a system at zero temperature and of finite size, referred to as the central region $C$, in contact with two macroscopic reservoirs (see Fig.~\ref{fig:SchemeEmil}). The central region consists of a finite square lattice with $N = N_x \times N_y$ sites, and at each lattice site there is one electronic orbital and one localized spin. The electronic orbitals are populated by spinful itinerant electrons that can tunnel in and out of the reservoirs, also referred to as leads. The leads couple only to the electronic degrees of freedom, i.e. there are no localized spins in the reservoirs. Inside the leads the electrons are assumed to be non-interacting and spin polarized, with a density set by the chemical potential. Also, in the central region the electrons are assumed to be non-interacting, although electron-electron interactions are straightforward (but numerically demanding) to include~\cite{LeadsPaper}. However, the itinerant electrons of the central region interact with the localized magnetic moments via a local exchange coupling, and the magnetic moments in turn interact among themselves via various magnetic interactions. The Hamiltonian of the composite system is
\begin{align}
H(t) = H_C(t) + H_R (t)+ H_{CR}, \label{H_tutto}
\end{align}
where a possible time-dependence is explicitly indicated
both for the central region and the reservoirs. This is necessary to initiate the dynamics of the system. We now discuss each contribution to $H$ separately.


\subsection{The central region}
The Hamiltonian of the central region is $H_C(t) = H_e + H_s + H_{s-e}$, where $H_e$ describes the itinerant electrons, $H_s$ the localized spins, and $H_{s-e}$ the spin-electron coupling. These terms are respectively given by
\begin{widetext}
\begin{align}
 H_e &= \sum_{\ev{ij}\sigma\sigma'} c_{i \sigma}^\dagger (-t_{ij} \mathbf{1} + \bm{\alpha}_{ij} \cdot \bm{\tau})_{\sigma \sigma'} c_{j\sigma'} - B \sum_i \hat{s}_{i}^z \label{Ham_e} \\
 H_s &= -\frac{1}{2} \sum_{\ev{ij}}J_{ij} \hat{\vb{S}}_i \cdot \hat{\vb{S}}_j - \frac{1}{2} \sum_{\ev{ij}} \vb{D}_{ij} \cdot \qty( \hat{\vb{S}}_i \cross \hat{\vb{S}}_j ) 
 - B \sum_i \hat{S}_{i}^z - \frac{K}{2} \sum_i (\hat{S}_{i}^z)^2 \label{Ham_S} \\
 H_{s-e} &= -g \sum_i \hat{\vb{S}}_i \cdot \hat{\vb{s}}_i. \label{Ham_eS}
\end{align}
\end{widetext}
Here $c^\dagger_{i\sigma}$ creates an itinerant electron at site $i$ with spin projection $\sigma$, the hopping amplitude between nearest-neighbor sites $i$ and $j$ (denoted by $\ev{ij}$) is given by $t_{ij}$, $\boldsymbol\alpha_{ij}$ accounts for spin-orbit interactions, and ${\bf B} = B \hat{\bf z}$ is an external magnetic field along the $z$-axis. The electronic spin operator at site $i$ is defined by $\hat{\bf s}_i = \sum_{\sigma\sigma'}c_{i\sigma}^\dagger \boldsymbol\tau_{\sigma\sigma'} c_{i\sigma'}$, where $\boldsymbol\tau$ denotes the vector of Pauli matrices. The parameters $J_{ij} = J_{ji}$ and ${\bf D}_{ij} = -{\bf D}_{ji}$ give the exchange and Dzyaloshinskii-Moriya interaction (DMI)~\cite{Dzyaloshinskii58,Moriya60} between spins $\hat{\bf S}_i$ and $\hat{\bf S}_j$, and $K$ quantifies an easy-axis single-ion anisotropy. The itinerant electrons interact with localized spins via a local exchange coupling of strength $g$.
We note that in magnetic thin films, skyrmions are often stabilized by the easy-axis anisotropy~\cite{Heinze11}, and an external magnetic field is not strictly necessary. However, including a Zeeman term in the Hamiltonian provides an additional physical means to control the equilibrium magnetic state, and in particular to tune the system between a ferromagnetic state and the skyrmion crystal phase. 

Depending on the symmetries of the system, the DMI vector ${\bf D}_{ij}$ can be of different form. Assuming the overall magnitude of the interaction is fixed at $|{\bf D}_{ij}| = D$, the spatial dependence typically takes one of the following two forms
\begin{align}
\vb{D}_{ij} &= D\, \vb{d}_{ij}  \hspace*{1.7cm} \text{(Bloch)} \\
 \vb{D}_{ij} &= D\, \vb{d}_{ij} \times \hat{\vb{z}}. \hspace*{1cm} \text{(N\'eel)} \nonumber
\end{align}
Here $\vb{d}_{ij}$ is the vector between lattice sites $i$ and $j$, and $\hat{\vb{z}}$ is a normal vector to the two-dimensional lattice plane. The N\'eel-type DMI gives rise to N\'eel-type (hedgehog) skyrmions, and is typically generated by the inversion symmetry breaking induced at a surface. The N\'eel type DMI is therefore most common in single layer or few layer substrates, since thicker materials tend to restore the bulk inversion symmetry. 
On the other hand, the Bloch-type DMI that commonly arises in systems with non-centrosymmetric crystal structures, gives rise to Bloch-type (spiral) skyrmions. Even so, many bulk materials that support Bloch-type skyrmions can be fabricated as thin film samples, while still hosting skyrmions~\cite{Experimental_review}.

In this work, we will limit ourselves to a Bloch-type DMI, while considering for computational simplicity a monolayer geometry for the localized spins. Similarly we will focus on a Rashba-type spin-orbit coupling of the form
\begin{align}
 \bm{\alpha}_{ij} = i\alpha_R\,\vb{d}_{ij} \times \hat{\vb{z}},
\end{align}
where the parameter $\alpha_R$ sets the overall magnitude of the interaction. This reflects the fact that the forms of the spin-orbit coupling and Dzyaloshinskii-Moriya interaction are intimately related, such that to lowest order in a strong coupling expansion ${\bf D}_{ij} \propto \boldsymbol{\alpha}_{ij}$~\cite{Moriya60}.
We note that the Hamiltonian $H_C$ can be generalized further by including both a local on-site potential $v_i(t)$ for the itinerant electrons, as well as a complex hopping $t_{ij} e^{i\phi_{ij}}$ whose phase encodes the interaction with an external electromagnetic field~\cite{LaserPaper}. It is similarly straightforward to include a direct coupling between the localized spins and external electromagnetic fields e.g. via the inverse Faraday effect. For the present work such terms are not of relevance, and are therefore omitted.


\subsection{The reservoirs and their coupling to the central region}\label{sec:reservoirs}
The leads are taken to be two-dimensional and semi-infinite (see Fig.~\ref{fig:Scheme}), and are described by the Hamiltonian 
\begin{align}
 H_R (t) &= \sum_{\alpha\sigma}H^R_{\alpha,\sigma} (t) \label{Hlead_1} \\
 H^{R}_{\alpha\sigma}(t) &= t_{R\alpha} \sum_{\ev{ij}\sigma} a_{i\sigma,\alpha}^\dagger a_{j\sigma,\alpha} + \sum_{i\sigma} v^{\alpha}_{\sigma}(t) a_{i\sigma,\alpha}^\dagger a_{i\sigma,\alpha}.
\end{align}
Here $\alpha \in \{L, R\}$ denotes the left (L) and right (R) lead, $\sigma$ labels the spin projection, and the parameter $t_{R\alpha}$ is the hopping amplitude within lead $\alpha$. The hopping amplitude is related to the nominal band width by $W_\alpha = 4t_{R\alpha}$, and $v^{\alpha}_{\sigma}(t)$ is a time-dependent bias measured from the chemical potential $\mu$. 
The leads interact with the central region via the Hamiltonian $H_{CR} = \sum_{\alpha\sigma} H^{CR}_{\alpha,\sigma}$, where
\begin{align}
& H^{CR}_{L\sigma}  = t_L \sum_{i_y=1}^{N_y} \big[ a^\dagger_{(1, i_y) \sigma, L} c_{(1, i_y )\sigma} + H.c. \big] \label{Hlead_2} \\
& H^{CR}_{R\sigma} = t_R \sum_{i_y=1}^{N_y} \big[ a^\dagger_{(1, i_y) \sigma, L} c_{(N_x, i_y )\sigma} + H.c. \big]. \label{Hlead_3}
\end{align}
Here $i_y$ denotes the $y$-component of the site index $i$, and runs over the transverse dimension $N_y$ of the system. For the left (right) lead, the values of the $x$-component $i_x$ run backward (forward) along the $x$-axis. The geometry of the system-reservoir system is
is pictorially illustrated in Fig.~\ref{fig:Scheme}, with semi-infinite and spin-polarized leads coupled to the left and right edges of the central system, supporting a spin-polarized current that interacts with the magnetic texture. Furthermore, a schematic of the 
parameters relevant to the transport setup, and to support charge and spin currents, is reported
in Fig.~\ref{fig:SchemeEmil}.


\section{Spin equations of motion}\label{sec:eom}
We now consider the system's nonequilibrium dynamics, which is initiated by applying a voltage bias to the reservoirs. Since the radius of a typical skyrmion is $\sim 100$ nm~\cite{Wang2018}, a full quantum mechanical description of the spin texture is extremely challenging. A commonly adopted strategy is then to resort to a classical description of the spins, where $\hat{\bf S}_i \rightarrow  \langle \hat{\bf S}_i \rangle \equiv {\bf S}_i = S {\bf n}_i$ and $|{\bf n}_i|=1$. This limit is exact for $S \to \infty$~\cite{Lieb73,Fradkin13}, and considered to be a suitable approximation already when $S > 1$~\cite{Heinze11}. Taking the classical limit leads to a semi-classical approximation for the coupled spin and electron subsystems~\cite{LeadsPaper,LaserPaper}, where the quantum electronic Hamiltonian depends parametrically on the classical variables ${\bf n}_i$. In this approximation the system's time evolution is therefore governed by two coupled differential equations, one for the classical spins and one for the quantum mechanical electrons. 


To obtain a dynamical equation for the localized spins, we start from the Heisenberg equations of motion for the spin operators and then take the classical limit. The result is a Landau-Lifshitz equation of the form
\begin{widetext}
\begin{align}\label{eq:spin_force} 
 \dv{\vb{n}_i}{t} &= -\vb{n}_i \cross \Big[ \sum_{\ev{j}} S(J_{ij}\vb{n}_j + \vb{D}_{ij}\cross \vb{n}_j) + (B + SK n_{iz}) \hat{\bf z} + g \langle\hat{\vb{s}}_i\rangle \Big],
\end{align}
\end{widetext}
where the last term gives the coupling between the classical spins and the instantaneous quantum average of the itinerant electron spins. In the following we absorb the factors of $S$ into the couplings, i.e. we define $S J_{ij} \rightarrow J_{ij}$ and similarly for $\vb{D}_{ij}$ and $K$, and for simplicity we assume $J_{ij} = J$. As commonly done in spintronics simulations, we add a small Gilbert damping to the equations of motion to stabilize the dynamics, which amount to adding an extra term $\alpha \vb{n}_i \cross (\partial \vb{n}_i/\partial t)$ in Eq.~(\ref{eq:spin_force}). The equations of motion for the itinerant electrons is discussed in detail in the next Section.
\section{Nonequilibrium Green's functions theory}\label{sec:negf}
To describe the dynamics of the electrons, we use the theory of non-equilibrium Green’s functions (NEGF) within the generalized Kadanoff-Baym ansatz (GKBA). NEGF are a general and powerful approach to nonequilibrium phenomena \cite{KadanoffBaym,Keldysh,balzer2012book,Stefanucci2013,Hopjan2014}, which describes the real-time dynamics of a system by exactly including external perturbations with an arbitrary temporal dependence. 
Within this theory the time-dependent expectation values of any single-particle observable, such as currents, densities and magnetization, can be obtained from the one-particle, two-time Green's function
\begin{align}\label{eq:greens_function}
  G_{ij}^{\sigma \sigma'}(z,z') = -i\ev{  \mathcal{T}_\gamma\qty{ c_{i\sigma}^H(z)[c^H_{j\sigma'}(z')]^\dagger } }_0.
\end{align}
Here the brackets $\ev{\cdot}_0$ denote an ensemble average with respect to the thermal density operator $\rho = e^{-\beta H}/Z$, where $\beta = 1/k_BT$ is the inverse temperature, $H$ the equilibrium Hamiltonian and $Z = \tr e^{-\beta H}$ the canonical partition function. In the low-temperature limit $\beta \to \infty$, the ensemble average reduces to a ground state expectation value. The operator $c_{i\sigma}^H(z)$ is the annihilation operator of an itinerant electron written in the Heisenberg picture with respect to the full time-dependent Hamiltonian $H(z)$, $z$ is a complex time argument living on the Keldysh contour \cite{Keldysh}, and all operators inside the brackets are time-ordered on the contour by the operator $\mathcal{T}_\gamma$. As before, the indices $i$ and $j$ run over all lattice sites, while $\sigma$ and $\sigma'$ denote spin projections.

We note that within the semi-classical scheme discussed above, the one-particle Green's function depends parametrically on the classical spin variables via the spin-electron coupling $H_{s-e}$. We should therefore write $G_{ij}^{\sigma \sigma'}(z,z') = G_{ij}^{\sigma \sigma'}(z,z'; \{{\bf S}_i\})$, but for notational simplicity we keep the dependence on ${\bf S}_i$ implicit in the following. In addition, we suppress the site and spin indexes of the Green's function, with the implicit understanding that all quantities are matrices in site and spin space, and only explicitly indicate the time variables of $G(z,z')$.
The equations of motion for the one-particle Green's function, the so-called Kadanoff-Baym equations, can be written as~\cite{KadanoffBaym}
\begin{align}\label{eq:kadanoff_baym}
 \qty[i \dv{z} -h(z) ]G(z,z') = \delta(z,z') + \int_\gamma \dd \bar{z}\, \Sigma(z,\bar{z}) G(\bar{z},z').
\end{align}
Here $h(z)$ is the time-dependent mean-field Hamiltonian, and the self-energy $\Sigma(z,z')$ contain all correlation effects beyond the Hartree approximation. A strength of the NEGF formalism is that non-interacting leads can be exactly incorporated in the equation of motion for the Green's function of the central system, through the introduction of a so-called embedding self-energy~\cite{Myohanen_2008}. Denoting the correlation and embedding self-energies respectively by $\Sigma_c$ and $\Sigma_{\rm emb}$, the total self-energy can be written as $\Sigma = \Sigma_c + \Sigma_{\rm emb}$. We note that in the present case the electronic Hamiltonian contains no interaction terms, and therefore $\Sigma_c = 0$. 

In the following we consider the wide-band limit (WBL) approximation to the embedding self-energy, obtained when the hopping amplitude inside the leads tend to infinity while the ratio  $t_{R\alpha}^2/t_{\alpha}$ remains fixed. In the formal treatment, this approximation is only performed in the extended direction of the leads (the one perpendicular to the system edge). In physical terms, it amounts the assumption that the density of states of the leads is constant over the band width of the central system.  Performing the WBL approximation results in the following expression for the embedding self-energy~\cite{Latini14,Tuovinen14,Ridley17,Tuovinen23}
\begin{align}\label{eq:sigma_wbl}
 \Sigma_{\rm emb}^<(t,t') &= i\sum_\alpha \Gamma_\alpha s(t)s(t') e^{-i\int_{t}^{t'}V_\alpha(\bar{t})\dd \bar{t}} 
  \int \frac{\dd\epsilon }{2\pi}f(\epsilon -\mu) e^{-i\epsilon(t-t')} \nonumber \\
 \Sigma_{\rm emb}^R(t,t') &= -\frac{i}{2}\delta(t-t') s(t)s(t')\sum_\alpha \Gamma_\alpha, \nonumber
\end{align}
where the superscripts denote the so-called lesser ($<$) and retarded ($R$) components of the self-energy \cite{Stefanucci2013}. Here $s(t)$ is a smooth function used to equilibrate the central system in presence of the leads, $V_\alpha(t)$ is the time-dependent bias in lead $\alpha$, and $f(\epsilon)$ is the Fermi-Dirac distribution.

\subsection{Generalized Kadanoff-Baym ansatz}
Provided that we have full knowledge of the self-energy $\Sigma$, the Kadanoff-Baym equations are an exact reformulation of the many-body problem. In practice however, $\Sigma$ needs to be approximated, which is commonly done using diagrammatic many-body perturbation theory. A computational difficulty met with the full Kadanoff-Baym formalism is that the numerical solution of Eq.~\ref{eq:kadanoff_baym} scales cubically with the number of time steps, which is a consequence of the memory integral (the right hand side of Eq.~\ref{eq:kadanoff_baym}) accounting for the full history of the dynamical evolution. Calculations with double-time Green's functions, as in Eq.~(\ref{eq:greens_function}), are highly expensive and scale unfavorably with basis size and simulation time. Since skyrmions typically occur on large lattice distances of $100$ nm~\cite{Nagaosa}, and move on typical time-scales of $1$ ps, the simulation of interacting spin-electron systems through a straightforward use of the Kadanoff-Baym equations is prohibitive. A significant improvement in the time-step scaling can be achieved by employing the so-called generalized Kadanoff-Baym ansatz (GKBA)~\cite{Lipavsky86}, where only the time-diagonal of the Green's function needs to be time evolved. This approximation was originally derived for equilibrium systems and in the weak scattering limit, but has been found to work well also out of equilibrium~\cite{Lipavsky86,Hermanns12,Hermanns14,Latini14,Bostrom18,Hopjan18,KarlssonHopjan18,Kalvova18,Joost20,Karlsson21,Pavlyukh22a,Pavlyukh22b,BonitzWBL23,Tuovinen23}. Within NEGF-GKBA and its recent reformulation as a time-linear scheme~\cite{Joost20,Karlsson21}, it is currently possible to simulate the long-time dynamics of electronic systems with a basis size on the order of $100$ orbitals~\cite{Pavlyukh22b}.

The GKBA for electronic degrees of freedom is achieved by the following factorization~\cite{Lipavsky86},
\begin{align}
 G^<(t,t') = iG^R(t,t') G^<(t',t')-iG^<(t,t) G^A(t,t'),
\end{align}
where the superscript $A$ denotes the advanced component of the Green's function. Using transformation rules introduced by Langreth ~\cite{Langreth1976}, we can project Eq.~\ref{eq:kadanoff_baym} from the Keldysh contour to the physical time axis, and obtain a dynamical equation for $G^<(t,t')$. When taking $t' \to t$ and using the GKBA, the equation for $G^<(t,t')$ reduces equation of motion for the one-particle density matrix $\rho(t) = iG^<(t,t)$ of the form.
\begin{widetext}
\begin{equation}\label{eq:gkba_eom}
 \frac{\partial\rho(t)}{\partial t} + i\comm{h(t)}{\rho(t)} = - \Big[ \int_{t_0}^t \dd\Bar{t}\, \Sigma^<(t,\Bar{t})  G^A(\Bar{t},t) 
 + \int_{t_0}^t \dd \Bar{t}\Sigma^R(t,\Bar{t})\rho(\bar{t})G^A(\bar{t},t) \Big] + H.c. 
\end{equation}
\end{widetext}
To close the equation for $\rho$ we further assume that $G^R(t,t') = -i\theta(t-t') \mathcal{T} e^{-i\int_{t'}^{t} d\bar{t}\, [h(\bar{t}) + i\Gamma/2]}$, with the interaction between the leads and the central region adiabatically turned on before the bias is applied. Here $\Gamma = \sum_\alpha \Gamma_\alpha$ accounts for the presence of the leads~\cite{Latini14}.


\section{Approximate wide band limit (AWBL)}\label{sec:awbl}
Introducing the GKBA improves the time-scaling of the Kadanoff-Baym equations, but solving the equation of motion for $\rho$ still scales quadratically with the number of time steps. Since we are interested in large systems with both fast and slow degrees of freedom, it is useful to make further approximations to reduce the time scaling. To this end, and inspired by other work on open systems ~\cite{Tuovinen14,Ridley17,BonitzWBL23,Tuovinen23}, we introduce here an approximate wide band limit (AWBL) based on an approximation to the collision integral (the right-hand side of Eq.~\ref{eq:gkba_eom}). As demonstrated below, this prescription offers a good trade-off between computation time and accuracy, and allows us to simulate large systems as required to describe skyrmion textures. 
The proposed AWBL amounts to neglecting the spin-electron coupling in the advanced Green's function $G^A$, such that
\begin{align}\label{mainAWBL}
 G^A(t,t') &= i\theta(t'-t) { \bar{\mathcal{T}}} e^{-i\int_{t'}^{t} d\bar{t}\, [H_e + H_{s-e}(\bar{t}) + i\Gamma/2]} \\
 &\approx i\theta(t'-t)e^{-i(t - t') (H_e + i\Gamma/2)}.
\end{align}
With this approximation the advanced Green's function becomes a function of the time difference, $G^A(t,t') = G^A(t - t')$. As a consequence, the entire collision integral becomes a function of $t - t'$, only a single evaluation needs to be performed at each time step, and the time-step scaling becomes linear (further technical details can be found in the Supplemental Material). 
In general the AWBL is a quite drastic approximation, but as demonstrated below it works rather well for the present system. One reason for this is that the spin-electron coupling only constitutes a higher-order correction to the embedding self-energy, since the spins do not couple directly to the leads. A second reason is that for most of the time-evolution, and in the dominant region of the central system, the spin texture is constant both in time and space.

\section{Skyrmion indicators}\label{sec:equilibrium}
Before discussing the response of magnetic skyrmions to electronic currents, we define the observables used to determine the presence of a magnetic skyrmion in a classical spin texture. Since the spins live on the unit sphere $S^2$, and the central region can be approximately compactified to the torus $T^2$ (assuming a ferromagnetic ordering along the edges), a topological charge measuring the winding number of the map ${\bf S}: T^2 \mapsto S^2$ can be defined. On a lattice, it can be shown that a suitable definition of this topological charge is given by~\citep{BERG1981}
\begin{align}
 Q &= \frac{1}{4\pi} \sum_{\{jkl\} \in \Delta} \Omega_{jkl},
\end{align}
where the sum is over the triangulated lattice and $\Delta$ is the set that contains the indexes for the lattice sites for every triangle in the lattice. The solid angle $\Omega_{jkl}$ is defined by
\begin{widetext}
\begin{align}
 \exp(i\Omega_{jkl}/2) &= \rho_{jkl}^{-1} \big(1 + \vb{S}_j \cdot \vb{S}_k + \vb{S}_k \cdot \vb{S}_l + \vb{S}_l \cdot \vb{S}_j \nonumber 
+ i \eta_{jkl} \vb{S}_j \cdot (\vb{S}_k \cross \vb{S}_l ) \big) \\
&\hspace{-6em} \rho_{jkl} = \sqrt{2(1+\vb{S}_j \cdot \vb{S}_k)(1 +\vb{S}_k \cdot \vb{S}_l) (1 +\vb{S}_l \cdot \vb{S}_j )}, 
\end{align}
\end{widetext}

and the function $\eta_{jkl}= {\rm sgn} [(\vb{S}_j \cdot (\vb{S}_k \cross \vb{S}_l )]$ ensures that the last term is positive. Geometrically the topological charge sums up the solid angles spanned by all spin triplets $\{jkl\}$, which when divided by the surface $4\pi$ of the unit sphere gives the integer winding number (for appropriate boundary conditions). The solid angle $\Omega_{jkl}$ serves as an indicator of the extent to which the spins twist. Specifically, within the core of a skyrmion, $\Omega_{jkl}$ has a large magnitude, gradually diminishing radially from the core. The extent of twisting at a specific location can be conceptualized as the skyrmion density at that particular site, with this density being invariably distributed across multiple sites. Furthermore, in the presence of a skyrmion within the system we define the center of mass as
\begin{align}\label{centermass}
 \vb{R}_{cm} &= \frac{1}{M}\sum_{i_x=2}^{N_x-1}\sum_{i_y=2}^{N_y-1} \rho_{i_x,i_y}  \begin{pmatrix} i_x \\ i_y \end{pmatrix},
\end{align}
The expressions for $M$ and $\rho_{i_x,i_y}$ are respectively given by
\begin{align}
M &= \sum_{i_x=2}^{N_x-1}\sum_{i_y=2}^{N_y-1} \rho_{i_x,i_y},\label{Mvalue}\\
\rho_{i_x,i_y} &= \bigg{|}\Omega_{(i_x,i_y),(i_x+1,i_y),(i_x+1,i_y+1)}\\
&+ \Omega_{(i_x,i_y),(i_x,i_y+1),(i_x+1,i_y+1)}\bigg{|}.
\end{align}
In Eqs.~(\ref{centermass}, \ref{Mvalue}), the index $i_x$ ($i_y$) ranges between 2 and $N_x-1$ (2 and $N_y-1$) rather than, as intuitively expected, between 1 and $N_x$ (1 and $N_y$). Observations of spin oscillations at the periphery of the central region motivate the exclusion of these edges in the determination of the skyrmion center of mass. The rationale behind these oscillations and the specific way we perform the site exclusion are further detailed in Section~\ref{bench}.


\section{Current driven skyrmion motion}\label{sec:current_motion}
We now consider the motion of skyrmions generated by a current density in the itinerant electron system, in turn driven by an external bias. However, first we need to prepare the system in a state featuring a skyrmion, a state that may not necessarily correspond to the ground state of the semi-classical system. This entails the selection of a suitable initial spin configuration, subsequently coupled in a self-consistent manner to the electron system. The interaction with the leads is slowly turned on in the time interval $t < \tau$ using the contact function $s(t) = \sin^2(\pi t/2\tau)$. After time $t = \tau$ a bias is applied, here considered to be of the form
\begin{align}\label{eq:bias}
 v_\alpha(t) = \theta(t-\tau) 
        \begin{cases}
          \phantom{-}V &\text{if } \alpha=L \text{ and } \sigma =\, \uparrow \\
         -V &\text{if } \alpha=R \text{ and } \sigma =\, \uparrow \\
          \phantom{-}0 & \text{otherwise}
        \end{cases}.
\end{align}
This bias generates a spin current through the central system.
In the simulations below we fix the the energy unit by setting the hopping $t_{ij} = t_s = -1$, and take $\hbar=1$.


\subsection{Benchmarks for the approximate wide band limit, and the role of damping}\label{bench}
\begin{figure*}
\begin{center}
\includegraphics[width=0.8\textwidth]{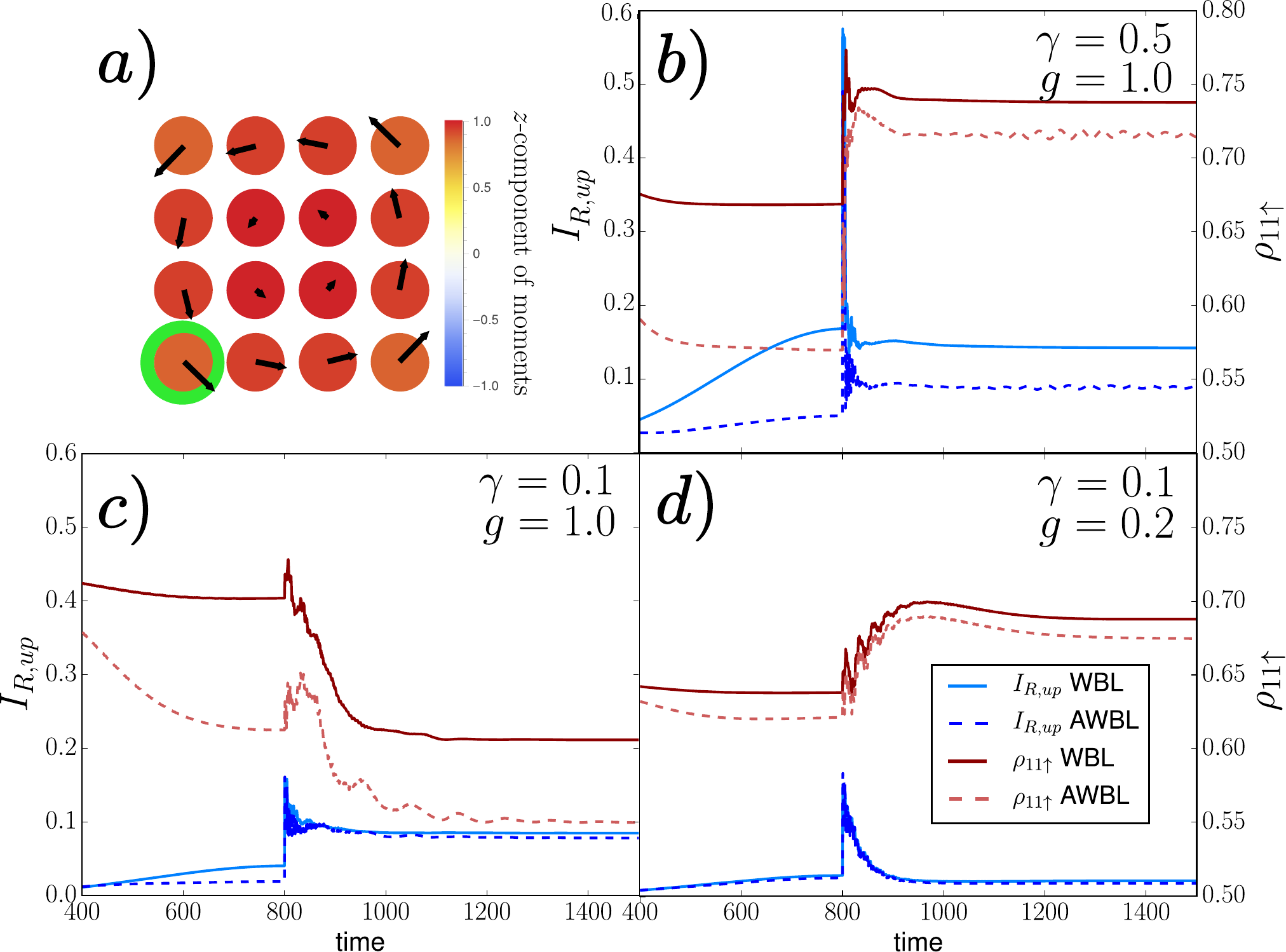}
\end{center}
 \caption{{\bf Benchmarks for the approximate wide band limit.} $(a)$ Initial relaxed configuration of a $4 \times 4$ spin texture interacting with a corresponding layer of electrons. $(b-d)$ Spin current $I_{R,\uparrow}$ and electronic spin density $\rho_{11\uparrow}$ of the lattice site encircled in green in $(a)$, for different values of $\gamma$ and $g$, as obtained within the wide band limit (WBL) and approximate wide band limit (AWBL). The spin texture is approximately the same for all cases and throughout the time evolution, due to a large value of the spin damping $\alpha$. The parameters used are $B = 0.05$, $K = 0.6$, $\alpha = 3$, $\alpha_R = 0$, $J = 0.5$, $D = 0.4$, $\tau = 800$, $\mu=0$ and $V = 3$. The color on the spins indicates their $z$-component, while the arrows show their direction in the $xy$-plane. The values of  $g$ and  $\gamma$ are given in each panel.}
 \label{fig:Comparison}
\end{figure*}

As a primer for our discussion of disorder effects on skyrmion dynamics, we briefly investigate the performance of the AWBL in the absence of disorder. For this purpose, we compare the dynamics obtained within the WBL and AWBL of a small central region denoted by $C_{16}$, consisting of a $4\times 4$ cluster as illustrated in Fig.~\ref{fig:Comparison}a.

We start from the ground state of the isolated region $C_{16}$, and gradually connect the system to the leads in the time interval $t\in [0, 800]$ measured in units of $\hbar/t_s$. This prepares the stationary initial state of the fully connected system. 
At time $t=800$, a symmetric and spin-polarized bias of the form shown in Eq.~\ref{eq:bias} is 
then applied. The subsequent part of the simulation, i.e. for times $t > 800$, is the one relevant for the skyrmion dynamics to be discussed later. To compare the performance of the WBL and AWBL, we choose as indicators the spin current and the spin-up density at a given site. Representative results are shown in Fig.~\ref{fig:Comparison}b-d, and indicate that the agreement between the WBL and AWBL varies noticeably with the strength of the spin-electron coupling $g$ and the lead-device connection $\gamma$. 

It is useful at this point to make a couple of remarks about the damping term
in the spin dynamics. The results in Fig.~\ref{fig:Comparison}b-d are obtained with a large Gilbert damping of $\alpha=3.0$, which results in a spin texture that remains essentially stationary during the time evolution. Therefore, these comparisons mainly concern the regime of electron dynamics in presence of a static magnetic background. Using the initial spin of Fig.~\ref{fig:Comparison}a) in $G^A$ instead of neglecting $H_{s-e}$ would give a very good agreement between the AWBL and the WBL. Strictly speaking, one can modify the AWBL to include a fixed spin texture in $G^A$ instead of neglecting it completely and still retain time-linear scaling. However, for the time-dependent disorder simulations to be discussed later, the spin texture changes significantly in time due to the skyrmion motion. We have verified numerically that, replacing in $G^A$ the exact contribution from the 
time evolving texture with a static one (chosen at any time during the time evolution) is not
better than using a $G^A$ with zero spin-electron coupling. This is why no spin configuration is included in $G^A$ in Eq.(\ref{mainAWBL}), or later on in the dynamics in presence of disorder.

\begin{figure*}[t]
\begin{center}
 \includegraphics[width=0.9\textwidth]{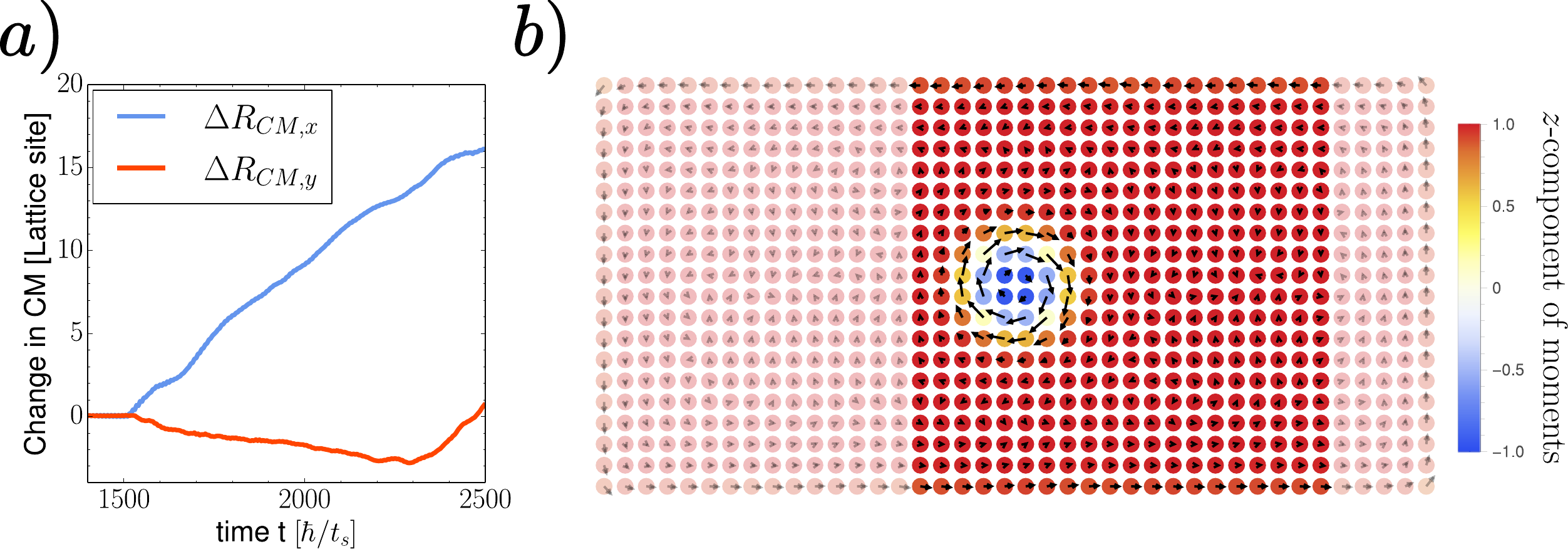}
 \end{center}
 \caption{{\bf Skyrmion motion in response to electronic spin currents.} (a) Center of mass motion of the skyrmion and (b) initial spin configuration at $t = 200$. The highlighted region corresponds to the area shown in Figs.~\ref{fig:case1} to \ref{fig:case3}. The parameters used are $g = 1$, $B = 0.05$, $K = 0.6$, $\alpha = 0.1$, $\alpha_R = 0$, $J = 0.5$, $D = 0.4$ (corresponding to Bloch-type DMI), $\gamma = 0.2$, $\tau = 1500$, $\mu = 0$ and $V = 2$. The color of the spins indicates their $z$-component, while the arrows display their direction in the $xy$-plane.}
 \label{fig:Base}
\end{figure*}

The purpose of using a large $\alpha$ in Fig.~\ref{fig:Comparison} is to attain a stationary initial state before the bias is applied. In Fig.~\ref{fig:Comparison}, such large damping is maintained throughout the whole time evolution (i.e. during the lead attachment and afterwards, when the bias is applied), for consistency and to focus on the electronic behavior. However, for the later results in the paper (Sect.~\ref{large_clean} and afterwards, where we consider 
skyrmion dynamics in central regions of $20\times40$ sites), 
the damping is set small ($\alpha=0.1$), in order not to influence the system's intrinsic dynamics, but still facilitate the adiabatic preparation of the initial state. In this respect, we have found that, even when evolving the system with no bias, and using very small or no damping after fully connecting the leads, there is a very slow build-up of oscillations in the current and the spin densities. Such oscillations primarily occur in the sites at the edges of the central region (and especially at the sites in contact with the leads). This holds
both for the small cluster in Fig.~\ref{fig:Comparison}a and for the larger systems studied later, and indicates that when the bias is applied the system has
has not yet reached in full the ground state in the presence of the leads, due to the very complex energy and fluctuating landscape provided by the interaction of the electrons with the classical spins. Though, in the spirit of having a microscopic current inducing the skyrmion motion, it is still meaningful to apply a bias to this configuration, and interpret the oscillations as physical in character. At the same time, these oscillations affect in a rather  artificial way the estimate of the skyrmion center of mass  $\vb{R}_{cm}$. Accordingly, in the results in the next sections, $\vb{R}_{cm}$ is
calculated using Eq.~\ref{centermass}, i.e  without including the contribution of the peripheral sites of the $20\times 40$ region.

Coming back to the results in Fig.~\ref{fig:Comparison}b-d, an interesting feature is that the agreement between the AWBL and WBL spin currents, for small $\gamma$, is generally better than agreement between the spin densities. While we do not have a clear explanation for this, we
note that this trend is confirmed in additional simulations (not shown), where $\gamma$ is varied at constant $g$ and viceversa. To summarize, our results suggest that the AWBL produces fairly accurate electronic and spin currents, overall in good agreement with the full WBL. It thereby constitutes a microscopic and semi-quantitative method to investigate skyrmion dynamics at low computational cost, appropriate for both transient and steady state regimes.

\begin{figure*}
 \centering
 \includegraphics[width=0.9\textwidth]{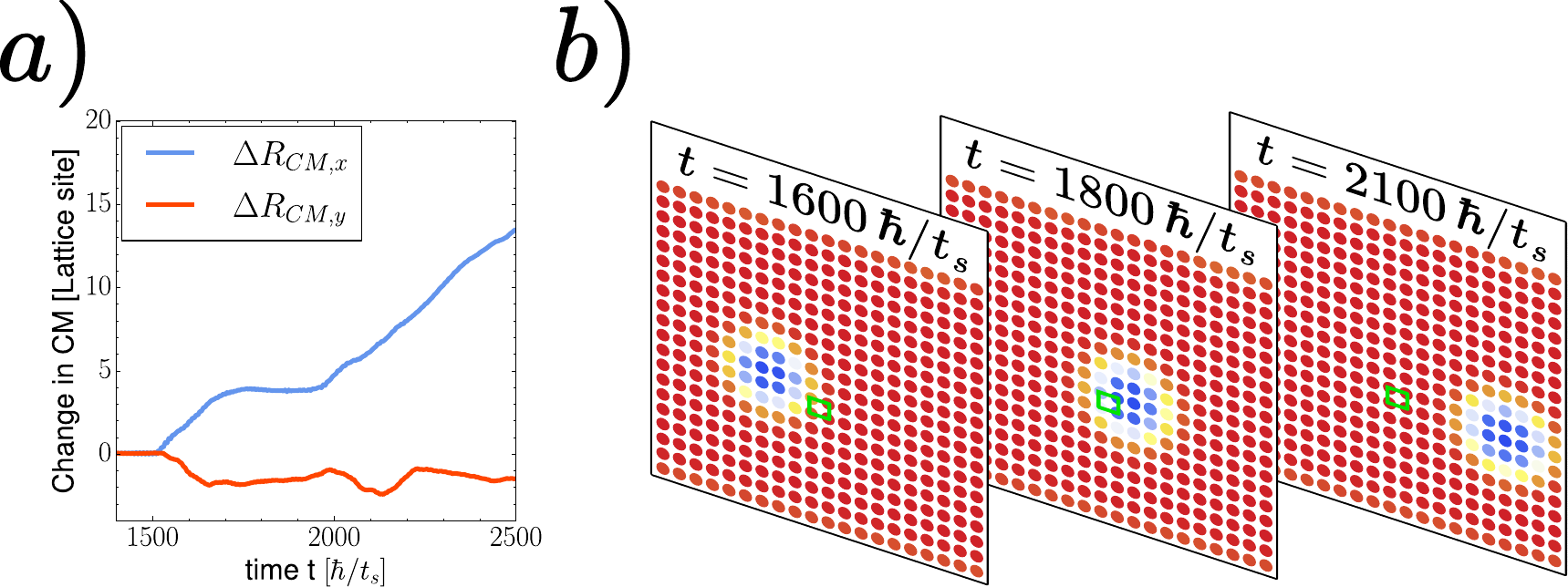}
 \caption{{\bf Skyrmion motion in presence of a small impurity on the skyrmion path.} (a) Center of mass movement of the skyrmion and (b) some snapshots of the $z$-component of the spins in the highlighted area of Fig.~\ref{fig:Base}. The parameters used are $g = 1$, $B = 0.05$, $K = 0.6$, $\alpha = 0.1$, $\alpha_R = 0$, $J = 0.5$, $D = 0.4$ (corresponding to Bloch-type DMI), $\gamma = 0.2$, $\tau = 1500$, $\mu = 0$ and $V = 2$. The color of the spins indicates their $z$-component.}
 \label{fig:case1}
\end{figure*}

\subsection{Current induced motion}\label{large_clean}
Having shown that the AWBL is in good quantitative agreement with the WBL for small systems, we now employ it to describe skyrmion motion in systems beyond the scope of the WBL. These results provide a useful benchmark for the discussion in later sections, the effect of magnetic impurities on skyrmion motion is analyzed. To obtain the reference results, we consider a central region with $40 \times 20$ sites, and investigate the dynamics in response to an applied bias. These results are henceforth referred to as "base case". In the base case as well as in the impurity studies below, a potential of strength $V = 2$ is applied in the leads, and the coupling between the system and leads is of strength $\gamma = 0.2$. 

The system starts off with a relaxed skyrmion texture in the center (see Fig.~\ref{fig:Base}b), and in response to the bias starts to drift towards the right with a slight additional downwards movement (see Fig.~\ref{fig:Base}a). As seen from the figure, the velocity in the $x$-direction is not constant  but rather ramps up, then slows down, and finally starts to ramp up again. While the dominant rightwards motion is due to the  spin current, the downwards force is due to the so-called skyrmion Hall effect~\cite{Nagaosa}.
Additionally, there are some small-scale oscillations, which are especially visible in the $y$-direction. These oscillations could be due to the discrete nature of the underlying lattice, since the skyrmion has a preferred equilibrium. When it moves between lattice sites, the skyrmion form gets distorted, but returns to its original configuration when it arrives at a new site.
We finally note that it takes some time for the skyrmion to pick up speed as the current is going through the system, which is a general trend in our simulations. This is indicative of the presence of a finite skyrmion mass, which has been discussed in previous studies \cite{PhysRevB.90.174434,PhysRevLett.109.217201} In particular, it was recently proposed that the magnitude of the skyrmion mass is affected by the interactions between localized moments and free electrons \cite{Nikolic1,Nikolic2}, consistent with the present findings.


\section{Disorder effects}\label{sec:disorder}
\begin{figure*}
 \centering
 \includegraphics[width=0.9\textwidth]{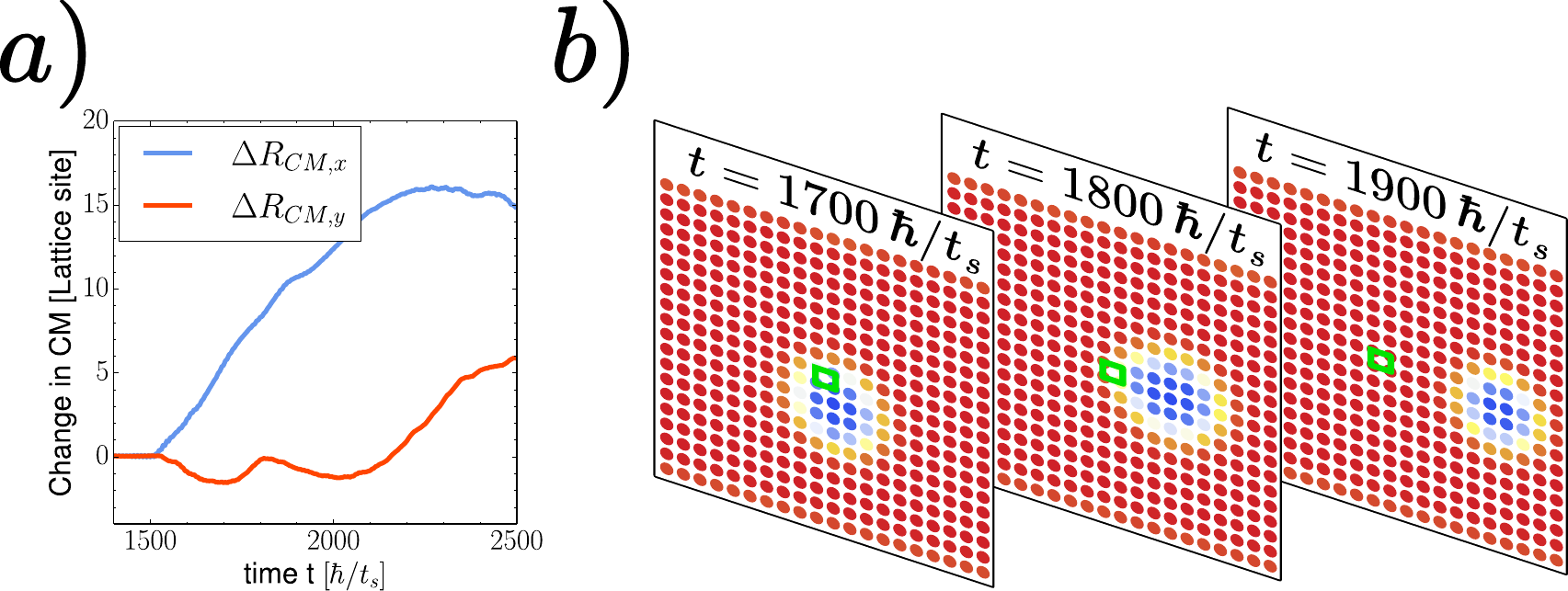}
 \caption{{\bf Skyrmion motion in presence of a small impurity near the skyrmion path.} (a) Center of mass movement of the skyrmion and (b) some snapshots of the $z$-component of the spins in the highlighted area of Fig.~\ref{fig:Base}. The parameters used are $g = 1$, $B = 0.05$, $K = 0.6$, $\alpha = 0.1$, $\alpha_R = 0$, $J = 0.5$, $D = 0.4$ (corresponding to Bloch-type DMI), $\gamma = 0.2$, $\tau = 1500$, $\mu = 0$ and $V = 2$. The color of the spins indicates their $z$-component.}
 \label{fig:case2}
\end{figure*}

Knowing the trajectory of the skyrmion in the base case, we can now investigate the effects of putting impurities in its path. In the panels of Figs.~\ref{fig:case1} to \ref{fig:case3} showing the spin configuration, the positions of the impurities are marked in green. Here we restrict to a simple form of impurity, corresponding to a reduction in the symmetric exchange interaction by $J \to J/2$ between the impurity site and its nearest neighbors. Therefore, it is expected that the skyrmion will be hindered by a smaller interaction. This change is effected at $t = 1400$ in order not to disturb the system before the bias is applied at $t = 1500$. 

\subsection{Case 1: A small impurity on the skyrmion path} 
As a typical example of a small magnetic impurity, we consider an impurity cluster consisting of a square of four lattice sites (see Fig.~\ref{fig:case1}). One possible realization of this form of disorder is the presence of adsorbate atoms in the center of these four lattice sites. The impurity is placed right in the path of the skyrmion, slightly to the right and below the skyrmion's initial position.

The center of mass movement of the skyrmion in response to a spin current is shown in Fig.~\ref{fig:case1}. In the $x$-direction, it seems like the skyrmion does not feel the effect of the impurity until it is on top of it, since the movement of the center of mass is quite similar to the base case before $t \sim 1700$. However, a careful comparison reveals that it moves slightly slower. In contrast, the skyrmion starts to pick up speed in the $y$-direction as it approaches the impurity, indicating that the velocity in the $x$-direction is transferred to a downwards motion.
After the skyrmion has entered the impurity, it gets pinned and struggles to escape. More specifically, between $t = 1700$ and $t = 2000$ the skyrmion is stuck to the impurity, and can only move around as long as some part of it is still on the impurity. A few times during the center of mass dynamics (see Fig.~\ref{fig:case1}), there are bumps, which indicate that the skyrmion gathered velocity to try to escape but was pulled back again by the impurity. Finally, at $t \sim 2000$, the skyrmion manages to escape and move past the impurity.
The dynamics presented in Fig.~\ref{fig:case1} clearly show the advantage of a microscopic description, where the motion of the skyrmion can be tracked atom by atom. During the dynamics the shape of the skyrmion is changing and the direction of motion changes several times, effects that are not possible to capture with a Thiele-type description. Furthermore, it is quite reasonable to expect that the observed dynamics are a direct result of the motion of the itinerant electrons, especially at the skyrmion boundary, where the local spin changes must take into account the redistribution of the electronic spin density during the time evolution.

\subsection{Case 2: A small impurity near the skyrmion path}
Next, we shift the impurity position two lattice sites upwards compared to the previous section (Fig.~\ref{fig:case2}), such that the impurity is now slightly above the skyrmion's path. At the beginning of the dynamics, the center of mass movement seems very similar to the previous case, with the velocity along the $x$-direction slightly reduced compared to the base case, and a significant downwards movement. However, in the present case the movement does not force the skyrmion on top of the impurity, and the impurity pinning is avoided. 
In the center of the mass movement of Fig.~\ref{fig:case2}, as well as in the snapshots at $t = 1700$ and $t = 1800$ of Fig.~\ref{fig:case2}, the skyrmion is seen to move beneath the impurity. It is likely that, if the system was extended further in the $y$-direction, the skyrmion would continue its motion downwards. In the present case however, the edge of the system prevents this, since our simulations suggest that the repulsive force induced by the edges makes the skyrmion to recoil upwards in the $y$-direction.
This illustrates, on the microscopic level, the importance of considering the track width when designing efficient racetrack memories. If the width of the track is too small, the skyrmion might not be able to get around impurities, while if the width is too large skyrmions can be deflected by impurities and substantially deviate from their path.

Although the skyrmion's core, shown in blue in Fig.~\ref{fig:Base}, is comprised of only $4 \times 4$ spins, a rotation of the $xy$-components of the spins away from the ferromagnetic background can be seen in an area extending up to $12 \times 12$ sites. This indicates that the effective skyrmion size is much larger than the central $4 \times 4$ region, which in turn implies that the skyrmion feels the presence of the edge much earlier that would be expected from only considering its core.


\subsection{Case 3: A larger impurity configuration}

\begin{figure*}
\begin{center}
 \includegraphics[width=0.9\textwidth]{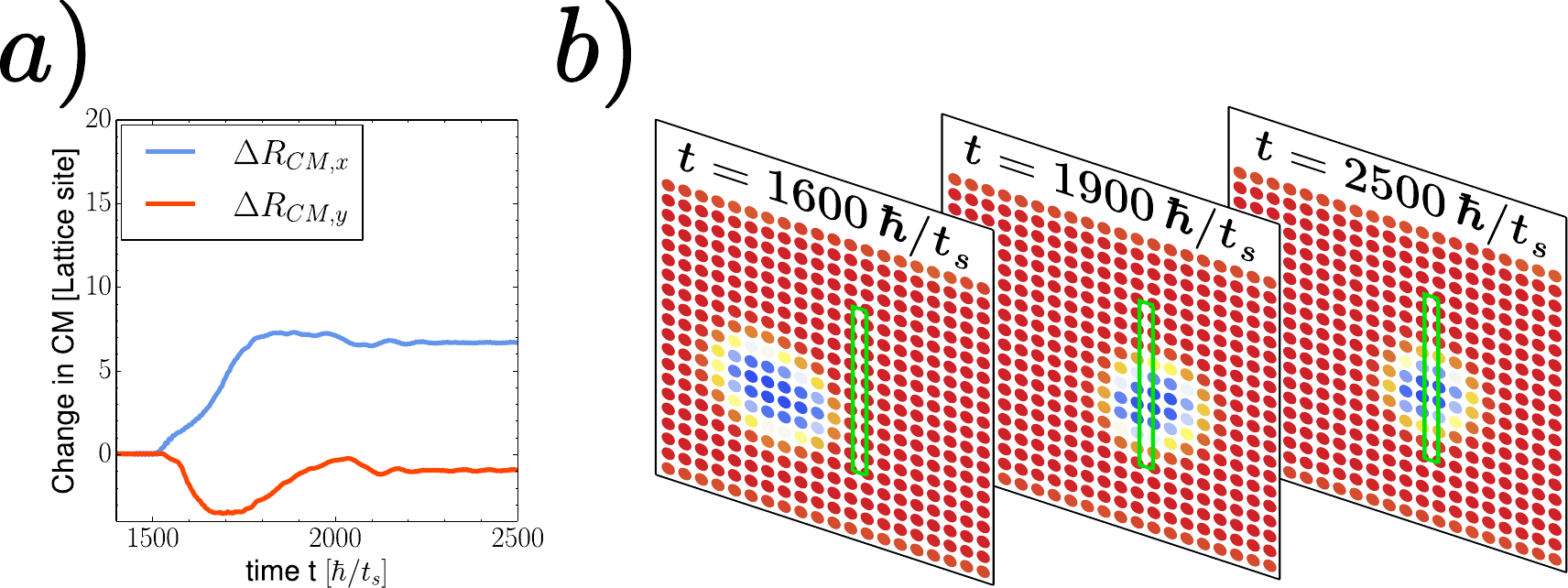}
\end{center}
 \caption{{\bf Skyrmion motion in presence of an extended impurity.} (a) Center of mass movement of the skyrmion and (b) some snapshots of the $z$-component of the spins in the highlighted area of Fig.~\ref{fig:Base}. The parameters used are $g = 1$, $B = 0.05$, $K = 0.6$, $\alpha = 0.1$, $\alpha_R = 0$, $J = 0.5$, $D = 0.4$ (corresponding to Bloch-type DMI), $\gamma = 0.2$, $\tau = 1500$, $\mu = 0$ and $V = 2$. The color of the spins indicates their $z$-component.}
 \label{fig:case3}
\end{figure*}

As an example of dynamics in presence of a more extended magnetic impurity, Fig.~\ref{fig:case3} shows the center of mass motion of the skyrmion in a system containing a columnar impurity that covers almost the entire lattice in the $y$-direction. Again, the impurity is only two lattice sites thick, but this time with only four sites at the top and the bottom left unperturbed. 
The characteristic downwards bump in the center of mass motion along the y-direction is also seen here, but it is twice as big as in the previous cases. As seen in Fig.~\ref{fig:case3}, the velocity in the $y$-direction is large before the skyrmion is suddenly halted. This detail, together with the observation that at $t = 1900$ the skyrmion is slightly smaller, suggests that a small skyrmion could be annihilated even by a small impurity. After $t = 1900$, the skyrmion recoils upwards but the collision with the top does not seem as severe as in the case of Fig.~\ref{fig:case1}. Likely, this is because the velocity is lower and the skyrmion is already stuck. 

The observed motion along the vertical direction is most likely aided by the lower wall. However, the impurity likely also plays an important role, since we have seen that it is hard for the skyrmion to escape once it moves on top of the impurity. In addition to the vertical motion, Fig.~\ref{fig:case3} also shows that the skyrmion keeps wiggling along the $x$-direction while moving. In particular, the snapshot at $t = 1900$ strongly indicates that this movement is due to the skyrmion expanding to the right. Subsequently, for times after $t \sim 2100$, the skyrmion shrinks and expands in a quasi steady state.


\subsection{Summary of the results}
To summarize this section, the results of Figs.~\ref{fig:case1} to \ref{fig:case3} show that by placing impurities in the selected places along a skyrmion's path, it is possible to engineer the skyrmion trajectory and to steer skyrmions in the desired direction in logic devices~\cite{PhysRevApplied.12.064053}. At the same time it is also clear that, in order to accurately engineer the desired skyrmion dynamics in the presence of impurities and provide some conceptual guidance for the experimental realization of skyrmion architectures, 
a microscopic and explicit equal-footing account of the behavior of itinerant electrons and localized spins is of high relevance.
\section{Discussion and outlook}\label{sec:discussion}
In this work we have presented a recently introduced method to deal with magnetic skyrmions, where a skyrmion magnetic texture made of classical spins is embedded in a background of quantum electrons. We considered the skyrmion dynamics in a quantum transport geometry, where a central region of coupled electrons and spins is connected to electron reservoirs. Our treatment is based on a two-component description where the electrons are treated via nonequilibrium Green's functions (NEGF), coupled to classical spins governed by a Landau-Lifshitz-Gilbert (LLG) equation. 

To describe the skyrmion motion in large systems, we introduced an approximate approach to include the effects electronic reservoirs coupled to the centeral system. This so-called approximate wide-band limit (AWBL) shows a satisfactory level of agreement with the full wide-band limit (WBL) approach. Similar to other schemes in the literature it is a time-linear scaling method, and in addition has a very advantageous scaling prefactor of great convenience to deal with systems of large size.

This approach was used to investigate the effects of magnetic impurities on the current-induced motion of magnetic skyrmion. Our results show that it is possible to characterize at the atomic level processes such as skyrmion scattering, recoil, drift, and trapping in disordered samples. This is a subject of high relevance, since understanding how tailor the disorder level of a given sample is crucial in order to control skyrmion dynamics. 

A natural extension of this work is to take into account the additional correlation effects arising from electron-electron interaction. This would bring the simulation closer to realistic systems, where correlations between the electrons could e.g. lead to destabilization of a skyrmion, or introduce effects a competing with disorder. In a different direction, the results obtained here can be used to benchmark other methods that deal with electron and localized spins. For example, one can imagine to validate  methods that, still in a framework, go beyond the standard LLG equation \cite{Nikolic1,Nikolic2}. 

Finally, we note that the methods presented here open the door for simulations of the intertwined spin-electron dynamics of small but realistic nanoscale systems. This provides a new approach to investigate the ultrafast dynamics of magnetic systems with general noncollinear orders, as initiated by ultrashort laser or current pulses~\cite{LaserPaper,Truc2023}. Apart from its importance for skyrmionic systems, the presented method will therefore be  of large relevance to describe the intertwined dynamics of spins, electrons and lattice vibrations in spintronics devices and for material engineering, thereby helping to facilitate a microscopic understanding of driven magnetic materials.


\section*{Conflict of Interest Statement}
The authors declare no competing financial or non-financial interests.\\

\section*{Author Contributions}
E.{\"O}. performed all calculations and interpretation of results
under the supervision of E.V.B. and C.V. E.{\"O} also contributed to
the extension of pre-existing code. The project was conceived by E.V.B. and C.V.
The overall supervision of the project was by C.V. All authors collaborated 
in writing the paper.

\section*{Funding}
EVB acknowledges funding from the European Union's Horizon Europe research and innovation programme under the Marie Skłodowska-Curie grant agreement No 101106809.
CV acknowledges funding from the Swedish Research Council (grant No VR 2022 04486)



\section*{Data Availability Statement}
The data supporting the findings of the present manuscript, as well as the files used to generate the figures, are available from the authors upon reasonable request.

\bibliography{bibliography2}


\clearpage

\end{document}


\appendix

\section*{Supplemental Material}



 There is one integral which needs to be evaluated in order to get the analytical expression for the lesser self-energy. Here we write $t$ instead of $t-t'$ for brevity.
 \begin{align*}
    I^<=\int_{-\infty}^{\infty}d\epsilon\, f(\epsilon-\mu)  e^{-i\epsilon t} =  \int_{-\infty}^{\infty}d\epsilon\, \frac{1}{1+e^{\beta (\epsilon-\mu)}}  e^{-i\epsilon t} =  \int_{-\infty}^{\infty}d\epsilon\, \frac{1}{1+e^{\beta (\epsilon-\mu)}}  e^{-i\epsilon t} e^{\eta \epsilon}
\end{align*}
Close contour in lower half plane since $t<0$, write $\Tilde{t}=t+i\eta$. The poles are at $\beta(\epsilon_n-\mu) = -i \pi - 2\pi i n$ with $n\in \mathbb{Z}$. If one has a finite hopping in the leads you get roughly $t_R \beta/\pi$ poles, since the integral has bounds $\pm 2t_R$ and $\beta$ makes the poles denser. This means that the sum below has a finite amount of terms and can therefore be well approximated by a finite amount of exponentials.
\begin{align*}
    I^< &= -2\pi i\sum_{n=0}^\infty e^{-i\epsilon_n \Tilde{t}}\text{Res}\qty[\frac{1}{1+e^{\beta (\epsilon_n-\mu)}}]  = 2\pi i/\beta \sum_{n=0}^\infty e^{-(\pi+i\mu\beta+2\pi n) \Tilde{t}/\beta}\\
    &=  2\pi i/\beta e^{-\pi \Tilde{t}/\beta-i\mu \Tilde{t}} \sum_{n=0}^\infty e^{-2\pi n \Tilde{t}/\beta} = 2\pi i/\beta e^{-\pi \Tilde{t}/\beta-i\mu \Tilde{t}} \frac{1}{1-e^{-2\pi \Tilde{t}/\beta}} = e^{-i\mu \Tilde{t}} \frac{i \pi/\beta }{\sinh(\pi \Tilde{t}/\beta)}
\end{align*}
\begin{align*}
    \text{Res}\qty[\frac{1}{1+e^{\beta (\epsilon_n-\mu)}}] &=  \lim_{\epsilon\to i/\beta(\pi+2\pi n)+\mu}\frac{(\epsilon-(\pi+2\pi n) i/\beta + \mu)}{1+e^{\beta (\epsilon-\mu)}} = \lim_{\epsilon\to i/\beta(\pi+2\pi n)}\frac{(\epsilon-(\pi+2\pi n) i/\beta)}{1+e^{\beta \epsilon}}\\
    &= \lim_{x\to 0}\frac{i/\beta x}{1+e^{i(x+\pi)}} =  \lim_{x\to 0}\frac{i/\beta x}{1- \qty(1+(ix)+\mathcal{O}\qty((ix)^2) )} = -1/\beta
\end{align*}
To generalize this result we consider the behavior for $t$ close to 0.
\begin{align}
 e^{-i\mu \Tilde{t}}\frac{i \pi/\beta }{\sinh(\pi \Tilde{t}/\beta)} = e^{-i\mu \Tilde{t}} \frac{i \pi/\beta }{\pi \Tilde{t}/\beta + \mathcal{O}\qty((\pi \Tilde{t}/\beta)^2)} \to \mathcal{P}\qty{\frac{i e^{-i\mu t}}{t}}  + \pi\delta(t)
\end{align}
Hence
\begin{align*}
    I^< = \mathcal{P}\qty{e^{-i\mu t} \frac{i \pi/\beta }{\sinh(\pi t/\beta)}} + \pi \delta(t)
\end{align*}
Since this is an integral coming from the lesser self-energy it is multiplied by $\theta(t)$, which means that in total we have:
\begin{align*}
    \mathcal{P}\qty{e^{-i\mu t} \frac{i \pi/\beta }{\sinh(\pi t/\beta)}} + \delta(t)\pi/2 
\end{align*}
This result has been obtained earlier in [\textit{Phys. Rev. B} \textbf{95}, 165440 (2017)].The low temperature limit is:
\begin{align*}
 \mathcal{P}\qty{\frac{i e^{-i\mu t} }{t} }
 + \delta(t) \pi/2 \quad \quad \quad \text{ when } {\beta \to \infty}
\end{align*}

\begin{align*}
    \int_{-\infty}^{\infty} d\epsilon\, \theta(t)e^{-i\epsilon t} &= \int_{-\infty}^{\infty} d\epsilon\qty(-\frac{1}{2\pi i}\int_{-\infty}^{\infty} d\epsilon'\,  \frac{1}{\epsilon' +i\eta}) e^{-i\epsilon' t}e^{-i\epsilon t} = -\frac{1}{2\pi i}\int_{-\infty}^{\infty} d\epsilon\int_{-\infty}^{\infty} d\epsilon'\,  e^{-i\epsilon t} \frac{1}{\epsilon' +i\eta} e^{-i\epsilon' t}\\
    &= -\frac{1}{2\pi i}  \int_{-\infty}^{\infty} d\epsilon \qty(-i\pi e^{-i\epsilon t}+\mathcal{P}\int_{-\infty}^{\infty} d\epsilon' \qty(\frac{1}{\epsilon'})e^{-i(\epsilon'+\epsilon) t})\\
    &= \frac{1}{2}  \int_{-\infty}^{\infty} d\epsilon e^{i\epsilon t} -\frac{1}{2\pi i} \int_{-\infty}^{\infty} d\epsilon \, \mathcal{P}\qty{ \int_{-\infty}^{\infty} dx \frac{e^{-ixt}}{x-\epsilon} }= \pi\delta(t)
\end{align*}
The second integral is the Hilbert transform $e^{-i x t}$ in x, if $t\neq 0$ then the Hilbert transform is $i\text{sgn}(t)e^{-i\epsilon t}$ which gives $\sim\delta(t)$ when integrated with $\epsilon$ and therefore it is zero. If $t=0$ we have the Hilbert transform of a constant, hence it's again zero. \\
The WBL matrix is given by
\begin{align}
 \qty[\Gamma^\alpha]^{\sigma\sigma'}_{ij} = \gamma  \delta_{\alpha_c i_x}\delta_{\alpha_c j_x }\frac{ \delta_{\alpha_s \sigma'}\delta_{\alpha_s \sigma } }{\pi}\sum_{k_y}  \frac{4}{N_y+1} \sin\left(\frac{\pi k_y i_y}{N_y+1}\right) \times \sin\left(\frac{\pi k_y j_y}{N_y+1}\right)
\end{align}
The alpha label contains spin $\alpha_s = \uparrow,\downarrow$ and coordinate $\alpha_c=1,N$ indices. $N$ is the rightmost site.\\
In the main text it is claimed that the following integral gives a time-linear evolution:
\begin{align}
  &\int_{t_0}^t \dd \Bar{t}\sum_\alpha e^{-i\int_{t}^{\Bar{t}}V_\alpha(\Tilde{t})\dd \Tilde{t}} \Lambda_\alpha(t-\Bar{t}) s(t)s(\bar{t})G^0(t-\Bar{t})\\
  &= s(t)\sum_\alpha\int_{t_0}^t \dd \Bar{t}  e^{-i\int_{t}^{\Bar{t}}V_\alpha(\Tilde{t})\dd \Tilde{t}} \Lambda_\alpha(t-\Bar{t}) s(\bar{t})G^0(t-\Bar{t}) = s(t)\sum_\alpha I_\alpha(t)
\end{align}
\begin{align}
  I_\alpha(t) &= \int_{t_0}^\tau \dd \Bar{t} e^{-i\int_{t}^{\Bar{t}}V_\alpha(\Tilde{t})\dd \Tilde{t}} \Lambda_\alpha(t-\Bar{t}) s(\bar{t})G^0(t-\Bar{t})+\int_{\tau}^t \dd \Bar{t} e^{-i\int_{t}^{\Bar{t}}V_\alpha(\Tilde{t})\dd \Tilde{t}} \Lambda_\alpha(t-\Bar{t}) s(\bar{t})G^0(t-\Bar{t})\\
  &=\int_{t_0}^\tau \dd \Bar{t} e^{-iV_\alpha(t-\bar{t})} \Lambda_\alpha(t-\Bar{t}) \sin(\pi\bar{t}/2 \tau)^2 G^0(t-\Bar{t})+\int_{\tau}^t \dd \Bar{t} e^{-iV_\alpha(t-\bar{t})} \Lambda_\alpha(t-\Bar{t}) G^0(t-\Bar{t}) \\
 &= I_{\alpha,1}(t)+I_{\alpha,2}(t)
\end{align}
\begin{align}
 I_{\alpha,1} = \int_{t_0}^\tau \dd \Bar{t} e^{-iV_\alpha(t-\bar{t})} \Lambda_\alpha(t-\Bar{t}) \qty(\frac{1}{2} - \frac{1}{4}e^{-i\pi\bar{t}/ \tau}- \frac{1}{4}e^{i\pi\bar{t}/ \tau}) G^0(t-\Bar{t}) = \frac{1}{2}I_{\alpha,1}^0-\frac{1}{4}I_{\alpha,1}^+ -\frac{1}{4}I_{\alpha,1}^-
\end{align}
\begin{align}
 I_{\alpha,1}^\Omega(t+\Delta) &= \int_{t_0}^\tau \dd \Bar{t} e^{-iV_\alpha(t+\Delta-\bar{t})} \Lambda_\alpha(t+\Delta-\Bar{t})e^{i\Omega\pi\bar{t}/ \tau}G^0(t+\Delta-\bar{t}) \\
 &=\int_{t_0-\Delta}^{\tau-\Delta} \dd y e^{-iV_\alpha(t-y)} \Lambda_\alpha(t-y)e^{i\Omega\pi(y+\Delta)/ \tau}G^0(t-y) \\
 &=\qty(\int_{t_0-\Delta}^{t_0}+\int_{t_0}^{\tau}-\int_{\tau-\Delta}^{\tau}) \dd y\, e^{-iV_\alpha(t-y)} \Lambda_\alpha(t-y)e^{i\Omega\pi(y+\Delta)/ \tau}G^0(t-y)\\
 &= e^{i\Omega \Delta/\tau}I_{\alpha,1}^\Omega(t)+\qty(\int_{t_0-\Delta}^{t_0}-\int_{\tau-\Delta}^{\tau}) \dd y\, e^{-iV_\alpha(t-y)} \Lambda_\alpha(t-y)e^{i\Omega\pi(y+\Delta)/ \tau}G^0(t-y)\\
 &\approx e^{i\Omega \Delta/\tau}I_{\alpha,1}^\Omega(t) + \frac{\Delta}{2}\Bigg{[}e^{-iV_\alpha(t-t_0+\Delta)} \Lambda_\alpha(t-t_0+\Delta)e^{i\Omega\pi(t_0)/ \tau}G^0(t-t_0+\Delta)\\
 &+ e^{-iV_\alpha(t-t_0)} \Lambda_\alpha(t-t_0)e^{i\Omega\pi(t_0+\Delta)/ \tau}G^0(t-t_0)\Bigg{]}\\
 &-\frac{\Delta}{2}\Bigg{[}e^{-iV_\alpha(t-\tau+\Delta)} \Lambda_\alpha(t-\tau+\Delta)e^{i\Omega\pi(\tau)/ \tau}G^0(t-\tau+\Delta))\\
 &+ e^{-iV_\alpha(t-\tau)} \Lambda_\alpha(t-\tau)e^{i\Omega\pi(\Delta+\tau)/ \tau}G^0(t-\tau)\Bigg{]}
\end{align}
So we need to evolve two GF:s, $G^0(t-t_0)$ and $G^0(t-\tau)$. The second one is only needed when $t>\tau$, since before then this integral is up to $t+\Delta$ and the change of variables on gives one new interval instead of two.
\begin{align}
 I_{\alpha,2}(t+\Delta) &= \int_{\tau}^{t+\Delta} \dd \Bar{t} e^{-iV_\alpha(t+\Delta-\bar{t})} \Lambda_\alpha(t+\Delta-\Bar{t}) G^0(t+\Delta-\Bar{t}) \\
 &=\int_{\tau-\Delta}^t \dd y e^{-iV_\alpha(t-y)} \Lambda_\alpha(t-y) G^0(t-y)\\
 &=I_{\alpha,2}(t)+\int_{\tau-\Delta}^\tau \dd y e^{-iV_\alpha(t-y)} \Lambda_\alpha(t-y) G^0(t-y)\\
 &\approx I_{\alpha,2}(t) + \frac{\Delta}{2}\bigg{[} e^{-iV_\alpha(t-\tau+\Delta)} \Lambda_\alpha(t-\tau+\Delta) G^0(t-\tau+\Delta)\\
 &+ e^{-iV_\alpha(t-\tau)} \Lambda_\alpha(t-\tau) G^0(t-\tau) \bigg{]}
\end{align}
This integral requires the GF $G^0(t-\tau)$ which we are already computing. \\